\documentclass[pra,twocolumn,superscriptaddress,showpacs]{revtex4}
\usepackage{amsmath,graphicx,amsfonts,bm,amssymb}
\usepackage{times}
\begin{document}

\title{Tunable interfaces for realizing universal quantum computation with topological qubits}

\author{Zheng-Yuan Xue} \email{zyxue@scnu.edu.cn}
\affiliation{Department of Physics and Center of Theoretical and Computational
Physics,\\ The University of Hong Kong, Pokfulam Road, Hong Kong, China}

\affiliation{Laboratory of Quantum Engineering and Quantum Materials, and School of
Physics and Telecommunication\\  Engineering, South China Normal University,
Guangzhou 510006, China}

\author{L. B. Shao}
\affiliation{National Laboratory of Solid State Microstructure and Department of Physics,
Nanjing University, Nanjing 210093, China}

\author{Yong Hu}
\affiliation{Department of Physics and Center of Theoretical and Computational
Physics,\\ The University of Hong Kong, Pokfulam Road, Hong Kong, China}
\affiliation{School of Physics, Huazhong University of Science and Technology,
Wuhan 430074, China}

\author{Shi-Liang Zhu}
\affiliation{National Laboratory of Solid State Microstructure and Department of Physics,
Nanjing University, Nanjing 210093, China}
\affiliation{Laboratory of Quantum Engineering and Quantum Materials, and School of
Physics and Telecommunication\\  Engineering, South China Normal University,
Guangzhou 510006, China}

\author{Z. D. Wang}\email{zwang@hku.hk}
\affiliation{Department of Physics and Center of Theoretical and Computational
Physics,\\ The University of Hong Kong, Pokfulam Road, Hong Kong, China}

\date{\today}

\begin{abstract}
We propose to implement tunable interfaces for realizing universal quantum computation with
topological qubits. One interface is between the topological and 
superconducting qubits, which can realize an arbitrary single-qubit gate on the topological qubit.
When two qubits are involved, the interface between the topological qubits and a microwave cavity
can induce a nontrivial two-qubit gate, which cannot be constructed based on braiding operations. The two interfaces, being tunable via an
external magnetic flux, may serve as the building blocks towards universal quantum computation with topological qubits.
\end{abstract}


\maketitle

%\section{Introduction}

Topological qubits are largely insensitive to local noises \cite{kitaev}, and
thus hold a promising future in quantum information processing. For  universal
quantum computation, one needs to encode a topological qubit with non-Abelian
anyons \cite{tqc}. Therefore, Majorana fermions (MFs) with non-Abelian
statistics have recently attracted strong renewed interest \cite{qi}. MFs are
a kind of self-conjugate quasiparticle  proposed in some systems
theoretically, e.g., certain vortex excitations in chiral \emph{p}-wave
superconductors \cite{tqc}. However, an unambiguous experimental verification
of MFs is still awaited. Recently, it is indicated theoretically that MFs can
be created on the interface between a strong topological insulator (TI)
\cite{ti} and an \emph{s}-wave superconductor by the proximity effect
\cite{fu}. Similar schemes with spin-obit coupling and \emph{s}-wave
pairing have also been proposed \cite{sau,alicea,zhu,wang,1d1,1d2,1d3}, which have
greatly advanced the field. Meanwhile, the interfaces between
topological qubits and quantum dots \cite{d1,d2,d3,xuedot}, as well as
superconducting qubits \cite{s1,s2,s3,xues1,jiang,zhangzt,xues2}, have also been
proposed. These hybrid systems may allow us to consolidate the
advantages of both types of qubits.

In this Brief Report, we propose tunable interfaces for realizing universal quantum computation
with topological qubits. Here, the hybrid system  is constructed with a
topological qubit, a superconducting charge qubit, and a microwave cavity. In
addition to an external magnetic flux, we also introduce a cavity-induced
magnetic flux in the superconducting qubit loop. In this way, an interface
between the topological qubit and the cavity, mediated by
the superconducting qubit, may be implemented. By modulating the external magnetic
flux, the interfaces between the topological qubit and the superconducting
qubit or the cavity can be switched on alternatively. For universal quantum
computation, the interface between the topological  and  the superconducting qubits is
sufficient for single-qubit control over and read out of the topological qubit, noting that
a topological qubit is usually hard to be read out. Another difficulty of
quantum computation with topological qubits lies in the fact that braiding can
not implement a nontrivial two-qubit gate. However, in our proposal,
when two qubits are involved, the interface between the topological qubits and a microwave cavity
can induce a nontrivial two-qubit gate. Therefore, the two interfaces may serve as
the building blocks towards universal quantum computation with topological qubits.

\begin{figure}[b]
\begin{center}
\includegraphics[width=8cm]{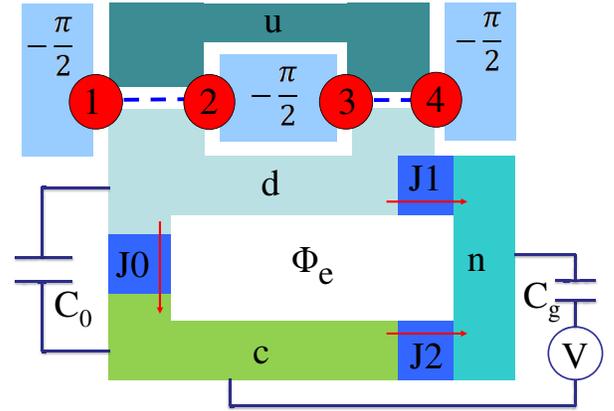}
\end{center}
\caption{(Color online) On the surface of TI, patterned
superconducting islands can form  hybrid system of topological and
superconducting qubits. Two pairs of MFs (red
circles) are localized at superconducting trijunctions,
connected by STIS quantum wires (dashed blue line). The coupling
between a pair of MFs is controlled by the superconducting
phases $\phi _{d}=\varepsilon$ and $\phi_{u}=-\pi$. The superconducting charge qubit consists of three JJs, enclosing
external magnetic fluxes from a dc magnetic field and a microwave
cavity field (not shown).  The superconducting phase $\phi_{c}$ is
fixed relative to $\phi_u$ by a phase-controller (not shown).
Then, $\phi_{d}$ will depend on the state of the superconducting
qubit and is related to the cavity field.} \label{qubit}
\end{figure}

We now proceed to introduce our considered setup. First, the topological qubit
is encoded by four MFs $\left\{\gamma_{i}\right\}  _{i=1,2,3,4}$, which satisfy the
fermionic anticommutation relation. A Dirac fermion  can be constructed from a
pair of MFs $\tilde{c}_{ij}^{\dag }=\left(\gamma_{i}-i\gamma_{j}\right)
/\sqrt{2}$, defining a twofold degenerated Hilbert space labeled by the
fermion parity $n_{ij}=\tilde{c}_{ij}^{\dag}\tilde{c} _{ij}=0,1$. In the even-parity 
subspace, a topological qubit is encoded with the basis states $|0\rangle
_{t}=|0\rangle_{12}|0\rangle_{34} $ and
$|1\rangle_{t}=|1\rangle_{12}|1\rangle_{34}$, where the subscript $t$ denotes
that the state is of the topological qubit. The MFs can be created on
the surface of a TI patterned with \emph{s}-wave superconductors \cite{fu}, and
thus the Cooper pairs can tunnel into the TI due to the proximity effect. Then,
assuming that the chemical potential is in the vicinity of  the Dirac point,
the Hamiltonian of the surface will obtain an additional \emph{s}-wave pairing
term. As shown in Fig. \ref{qubit}, each MF, indicated by a red circle, is
localized at  a point where three separated superconducting islands meet, i.e.,
a superconducting trijunction. The MFs can be created via a
superconductor-TI-superconductor (STIS) wire that separates the islands with
superconducting phases $\phi_{d}=\varepsilon$ and $\phi_{u}=-\pi$. A narrow
STIS wire (width $W\ll v_{F}/\Delta_{0}$) is  described by \cite{fu}
\begin{equation}
H_{STIS}=-iv_{F}\tilde{\tau}_{x}\partial_{x}+\delta_{\varepsilon}\tilde{\tau}_{z},
\end{equation}
where $v_{F}$ is the effective Fermi velocity, $\Delta_{0}$ is the
\emph{s}-wave superconducting  gap, $\delta_{\varepsilon}
=\Delta_{0}\cos\left[\left(\phi_{d}-\phi_{u}\right) /2\right]=-\Delta_{0}\sin
\varepsilon/2$, and $\tilde{\tau}_{x,z}$ are the Pauli matrices acting on the
wire's  zero modes. Figure \ref{qubit} shows two pairs of MFs with distance $L$, which
encode our topological qubit. The two pairs of
MFs  share the same type of coupling; e.g., for $\gamma_{1}$ and $\gamma_{2}$ the coupled Hamiltonian is
$\tilde{H}_{12}^{\mathrm{MF}}=iE\left(
\varepsilon\right)\gamma_{1}\gamma_{2}/2$ with an energy splitting
$E\left(\varepsilon\right)$ depending on the superconducting phase $\phi
_{d}=\varepsilon$. An effective Hamiltonian for the topological qubit is
\begin{equation}\label{mf}
H_{T}=-\frac{E\left(\varepsilon\right)}{2}\tau_z,
\end{equation}
where $\tau_z=\left(  \left\vert 0\right\rangle \left\langle 0\right\vert -\left\vert
1\right\rangle \left\langle 1\right\vert \right) _{\mathrm{t}}$ is the Pauli matrix acting on
the topological qubit and
\begin{equation}
E\left(  \varepsilon\right)  = {v_F \over L }\sqrt{\Lambda_{\varepsilon}^{2}
+f_{0}^{2}\left(  \Lambda_{\varepsilon}\right)  }
\end{equation}
with the dimensionless parameter $\Lambda_{\varepsilon}=\frac
{\Delta_{0}L}{v_{F}} \sin\frac{\varepsilon}{2}$ and $f_{n}\left( y\right)$
(with $n=0, 1, 2...$) being the inverse function of $y=x/\tan\left(  x\right) $
associated with the $n$th invertible domain. For $\Lambda_{\varepsilon} \gg1$
and $0<\varepsilon<\pi/2$, the topological qubit splitting
$E\left(\varepsilon\right)$ is negligibly small \cite{fu}. While for
$\Lambda_{\varepsilon}\leq 1$, $E\left(\varepsilon\right)$ becomes sensitive to
$\varepsilon$. To couple the topological and superconducting qubits, it is
natural to make $\varepsilon$ dependent on the superconducting qubit state. As
shown in Fig. \ref{qubit}, this can be achieved by making the superconductor
labeled "d" be a part of the superconducting qubit, and thus 
$\varepsilon$ is related to the magnetic flux pierced in the qubit loop.
Meanwhile, the superconducting qubit is placed in a cavity, and thus the
magnetic flux contains the external magnetic flux and the magnetic flux
comeing from the cavity. In this way, coupling among the three elements can be
implemented.

We now detail the coupling of the elements. As also shown in Fig. \ref{qubit}, the
superconducting charge qubit  \cite{vi,v2,you} consists of a small superconducting box
with $n$ excess Cooper-pair charges, formed by a symmetric superconducting quantum
interference device including two small identical Josephson junctions (JJs) with
capacitance $C_{J}$ and Josephson coupling energy $E_J$, and pierced by an external
magnetic flux $\Phi_e$. Meanwhile, a control gate voltage $V$ is applied via a gate capacitor $C_g$.
To slightly modulate the superconducting phase $\phi _{d}$, a
large JJ is also introduced, which has a Josephson coupling energy of $E_{J0} \gg  E_{J}$
and a capacitance of $C_{J0}$. In order to eliminate the influence of the charging energy of
the large JJ to the superconducting  charge qubit Hamiltonian,  a large capacitance $C_0$
is placed in parallel with  the large JJ \cite{vi}. Assuming that the inductance of the
qubit circuit is much smaller than that of the large JJ, the Hamiltonian of the superconducting qubit
can be written as  \cite{you}
\begin{eqnarray}\label{AA}
H_{S}=E_{c}( n-n_{g})^2-E_{J}(\cos \phi_{1}
+\cos \phi_{2})  -E_{J0}\cos \phi_J,
\end{eqnarray}
where $E_{c}= 2e^2/( C_g+2C_{J})$ is the charging energy of the superconducting island,
$n_{g}=C_gV /(2e)$ is the  induced charge of the gate voltage, and $\phi _{J}$, $\phi_{1}$
and $\phi_{2}$   are the phase drops across  JJs 0, 1 and 2, respectively.

Meanwhile, the superconducting charge qubit is placed at a magnetic antinode of
the cavity in a circuit QED scenario \cite{cqed}. For simplicity, we assume
that the cavity has only a single mode to play a role, the free Hamiltonian of
which is $H_c= \omega_r a^\dagger a$ (assuming $\hbar=1$ hereafter) with
$\omega_r$, $a$, and $a^\dagger$ being the frequency, annihilation, and creation
operators of the cavity mode, respectively. Flux quantization around the qubit
loop  leads to $ \phi_{1}= \phi-\beta$ and $ \phi_{2}= \phi+\beta$, where
$2\beta=\phi_e-\phi_J+2g(a+a^\dagger)$ with $\phi_e=2\pi\Phi_e/\Phi_0$ and $g$
is the magnetic coupling strength between the cavity and the superconducting
qubit; the average phase drop $\phi= (\phi_{1}+ \phi_{2})/2$ is canonically
conjugate to $n$   as $[ \phi, n]=i$. Consequently, the qubit Hamiltonian in
Eq. (\ref{AA}) can be rewritten as \cite{you}
\begin{eqnarray}\label{BB}
H_{CS}= E_{c}( n-n_{g})^2-2E_{J}\cos\beta\cos \phi -E_{J0}\cos\phi_J.
\end{eqnarray}

The induced circulating supercurrent in the qubit circuit is $I=2I_{c}\sin\beta\cos
\phi$, with $I_{c}=\pi E_{J}/\Phi_0$ being the critical current of the two small JJs.
Meanwhile, this supercurrent also flows through the large JJ, and thus $I=I_0\sin\phi_J$,
with $I_0=2\pi E_{J0}/\Phi_0$ being the critical current of the large JJ. Therefore,
\begin{equation}
I_0\sin\phi_J= 2I_{c}\sin\beta \cos \phi.
\label{A1}
\end{equation}
As $E_{J}\ll E_{J0}$,  $\phi_J$ will be small. Up to the second order of the
small parameter of $\eta= I_{c}/ I_0 $, we have
\begin{eqnarray}
\phi_J&=&2\eta\sin {\phi_e \over 2} \cos \phi -\eta^2\sin\phi_e\cos^2 \phi\notag\\
&&+2g\eta\cos {\phi_e \over 2} \cos \phi \times (a+a^\dagger).
\end{eqnarray}

At low temperatures ($k_BT\ll E_{c}$) and within the charging regime ($E_{J}
\ll E_{c} \ll \Delta_0$), only the lowest two charge states $\{|0\rangle_s,
|1\rangle_s\}$ are relevant for the superconducting qubit operating at its
degeneracy point ($n_g=1/2$), where the subscript $s$ denotes that the state is
of the superconducting qubit hereafter. As a result, the Hamiltonian in Eq.
(\ref{BB}) reduces to
\begin{equation}
H_{CSC}= -{ \overline{E}_{J} \over 2}  \sigma_x  +\xi (a+a^\dagger)\sigma_x,
\label{CC}
\end{equation}
where $\overline{E}_{J}=2E_{J} \cos {\phi_e \over 2} (1-{3\over 8}\eta^2 \sin^2
{\phi_e\over 2})$,  $\xi= gE_J\sin {\phi_e  \over 2}$, and $\sigma_{x,z}$ are
Pauli matrices acting on the superconducting qubit state. Meanwhile, in the
superconducting qubit representation, $\phi_J= f_1+(f_2+f_3)\sigma_x$, where
$f_1=-{1 \over 4} \eta^2\sin \phi_e  $, $f_2=\eta \sin{\phi_e \over 2}$, and
$f_3= \eta g (a+a^\dagger) \cos{\phi_e \over 2}$, which depends on the states
of the superconducting charge qubit and the cavity. If we fix $\phi_{c}$
\textrm{with respect to} $\phi_{u}$ with a phase controller, up to the second
order of $\eta$, $\phi_{d}$ will be  $\varepsilon^{+}=\phi_{c}+f_1+f_2+f_3$ and
$\varepsilon^{-}=\phi_{c}+f_1-f_2-f_3$, depending on the state of the
superconducting charge qubit in the states $|+\rangle_s$ and $|-\rangle_s$,
respectively. As $\eta$ is small, the separation  of $\phi_{d}$, $
\Delta\varepsilon=2(f_2+f_3)\propto \eta$, will be small, as we expected.

Finally, the combined hybrid system can be described by $H_{total}=   \omega_r a^\dagger
a -{1 \over 2}  \omega_t  \tau_z+ H_{CSC}+H_{int}$, with the interaction between the
topological qubit and others being
\begin{eqnarray}
H_{int}=-  {\lambda_1 \over 2} \sigma_x \tau_z -  \lambda_2   \sigma_x \tau_z (a+a^\dagger),
\label{hybrid}
\end{eqnarray}
where $\omega_t=E\left(\phi_c+f_1\right)$, $\lambda_1=  \eta \sin {\phi_e\over
2} {dE(\phi)\over d\phi}|_{\phi=\phi_c+f_1}$, and $\lambda_2=  \eta g \cos
{\phi_e\over 2} {dE(\phi)\over d\phi}|_{\phi=\phi_c+f_1}$. It is obvious that
$\lambda_{1,2}$ can be tuned via the external magnetic flux $\Phi_e$. In
particular, when $|\lambda_1|$ ($|\lambda_2|$) reaches its maximum value,
$\lambda_2$ ($\lambda_1$) will be  0. That is to say, we can selectively
implement  the topological and superconducting qubits interface or the
topological qubit and cavity interface. This is  distinctly different from the
proposed interface in Ref. \cite{jiang}, where the only  implemented interface
is between the topological and superconducting flux qubits.

We now consider the interface between  the topological and superconducting
charge qubits which can be switched on by modulating the external magnetic flux
to fulfill $\sin {\phi_e\over 2} =1$ ($\cos {\phi_e\over 2} =0$). With
$\lambda_1 t_1=-\pi/2$, up to local rotations on the superconducting qubit and
Hadamard gates on the topological qubit, we can implement arbitrary unitary
transformations for the two-qubit hybrid system \cite{jiang3,cirac,xue}.  For universal quantum
computation, this interface is
sufficient for single-qubit control over and read out of the topological qubit.

We now estimate the coupling strength of the interface with typical experimental
parameters. For the superconducting qubit, we may choose the following parameters  \cite{v2}:
$E_{J}=16$ GHz and $E_{J0}=10E_J$, which means $\eta=0.1$. For the topological qubit,
reasonable parameters are the following \cite{vi,jiang}: $\Delta_0=2\pi\times
32$ GHz, $v_F=10^5$ m/s, $L=5$ $\mu$m, and $T=20$ mK. Therefore, the maximum coupling
strength of this interface is $\lambda_1^{max}\approx\eta \Delta_0 =2\pi\times 3.2$  GHz.

The relevant imperfections of this interface are estimated as the following. First, as
$\lambda_1/(2E_J)=0.1$, the undesired tunneling probability between the qubit states
is suppressed to $P_t\sim0.01$. Second,  to suppress the  quantum fluctuations of the
large JJ, $C_0=100(C_g+2C_J)$ is chosen to make its effective charging energy negligible
small \cite{v2}, and thus it works in the classical regime \cite{mak}. Finally,
excitation of  the quantum wire modes with energy $E\approx v_F/L$  can be exponentially
suppressed to $P_e\sim \exp[-E/(K_B T)] < 10^{-3}$ \cite{jiang}.

We move to the topological qubit and microwave cavity interface by modulating
$\cos {\phi_e\over 2} =1$. The coupling between these two subsystems is
mediated by the superconducting charge qubit. Without loss of generality, we
assume that the superconducting qubit is initially prepared in its ground
state. We further tune the energy splitting of the superconducting charge qubit
far away from the cavity frequency so that the superconducting qubit will
always stay in its ground state. When two indentical qubits are involved,
the hybrid system is described by the interaction
\begin{eqnarray}
H_{CT}=\omega_r a^\dagger a-{1 \over 2}(\hbar \omega_t + \lambda_1) (\tau^z_1+\tau^z_2)\notag\\
-  \lambda_2  (\tau^z_1+\tau^z_2) (a+a^\dagger).
\label{tp1}
\end{eqnarray}
Setting $\phi_e'=2\omega t+\phi_e$, the interaction Hamiltonian in the interaction picture reads \cite{xue2}
\begin{eqnarray}
H_{I}=-  \lambda_2  \left( a e^{-i\nu t}+a^\dagger e^{i\nu t} \right) J_z,
\label{tp2}
\end{eqnarray}
where $\nu=\omega_r-\omega>0$ and $J_z=\sum_{j}\tau^z_j/2$.
The time-evolution operator for the Hamiltonian in Eq. (\ref{tp2}) can be expressed
as \cite{ms,zw}
\begin{eqnarray}\label{U0}
U(t) &=& \exp\left[{-iA(t)J _{z}^{2}}\right]\notag\\
&& \times \exp\left[{-iB(t)aJ_{z}}\right]\exp\left[{-i{B^{\ast}}(t)a^{\dagger}J_{z}}\right],
\end{eqnarray}
where
\begin{eqnarray}
A(t)=-\frac{\lambda_2 ^{2}}{\nu}\left[t - \frac{1}{i\nu}\left(
e^{i\nu t}-1\right)\right],
\end{eqnarray}
\begin{eqnarray}\label{b}
B(t)=-i\frac{\lambda_2 }{\nu}\left(e^{-i\nu t} -1 \right).
\end{eqnarray}
It is obvious that $B(t)$ is a periodic function of time
and vanishes at $\nu \tau=2k\pi$ where $ k=1,2,3,...$. At this time, the
operator in Eq. (\ref{U0}) reduces to
\begin{eqnarray}\label{u1}
U(\tau)=\exp\left[-i A(\tau) J_{z}^{2}\right],
\end{eqnarray}
with $A(\tau)=-\lambda_2 ^{2}\tau/\nu$. The maximum coupling strength of
this interface is $\lambda_2^{max}\approx\eta g \Delta_0=2\pi\times 32$  MHz
for $g=0.01$. In this way, we achieve the coupling between the topological
qubits mediated by the microwave cavity and the operator in Eq. (\ref{u1})
serves as a nontrivial two-qubit gate.

For example, choosing $A(\tau) =-\pi/2$ and the initial state of the two
topological qubits being $|\psi\rangle_i=|++\rangle_t$, the final state is
\begin{equation}\label{EPR}
|\psi\rangle_f=\frac{1}{\sqrt{2}}(|++\rangle_t+i|--\rangle_t),
\end{equation}
where $|\pm\rangle_t=(|0\rangle_t+|1\rangle_t)/\sqrt{2}$.
Note that $A(\tau) =-\pi/2$ can be achieved by choosing $\nu=2\lambda_2 \sqrt{k}$,
and thus the gate time is $\tau=\sqrt{k} \pi/\lambda_2$.
It is noted that the gate time will be increased for larger $k$, which leads to more severe
decoherence effect. Therefore,  $k=1$ is usually adopted.

\begin{figure}[tbp]
\includegraphics[width=8.5cm]{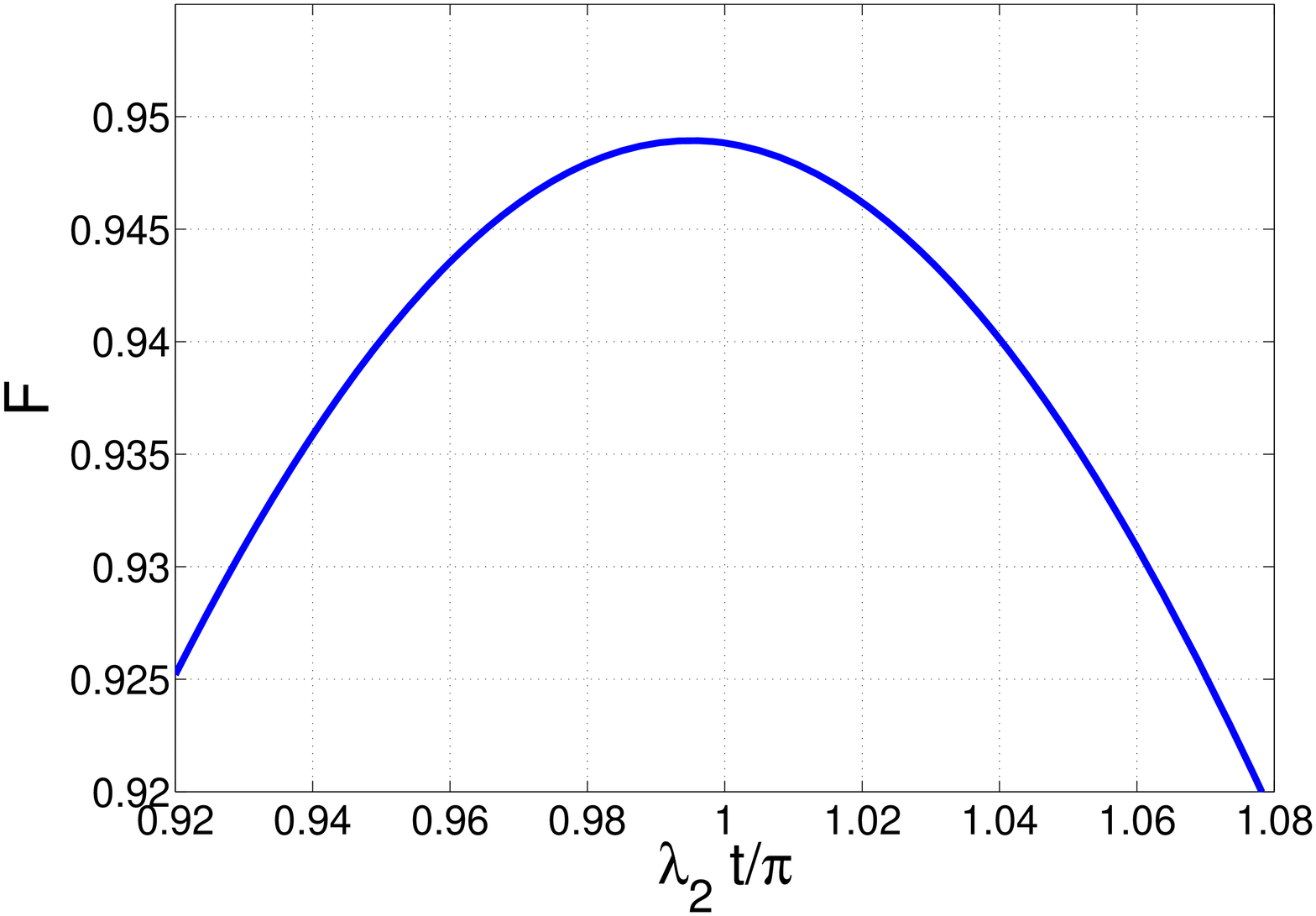} \label{f}
\caption{(Color online) The fidelity of entangling the two  topological qubits as a function of $\lambda_2 t/\pi$.
The parameters are $k=1$, $\kappa=\gamma=1$ MHz, and $\lambda_2=2 \pi
\times 32$ MHz. }
\end{figure}

We next consider the influence of dissipation to the entangling gate
by integrating the  quantum master equation,
\begin{eqnarray}  \label{me}
\dot{\rho}=-i [H_I, \rho] + \kappa (2a\rho a^\dagger -a^\dagger a \rho -\rho a^\dagger a)\notag\\
+\gamma \sum_{j=1}^2(2\tau_j^-\rho\tau_j^+ -\tau_j^+ \tau_j^- \rho -\rho \tau_j^+ \tau_j^-),
\end{eqnarray}
where $\rho$  is the density matrix of the combined system of the topological qubit and
cavity photon and $\kappa$  and $\gamma$ are the decay rate of the cavity, and the lifetime
of the topological qubit, respectively. We characterize the entanglement generation process
by the conditional fidelity of the quantum state defined by
$F=\langle\psi_f|\rho_a|\psi_f\rangle$, with $\rho_a$ being the reduced  density
matrix of the topological qubits. In Fig. 2, we plot the fidelity $F$ with $k=1$ as a function of dimensionless time
$\lambda_2 t/ \pi$, where we have obtained high fidelity $F\simeq 95\%$  for the generation.
In the plot, we have chosen the conservative parameters of  $\kappa= \gamma=1$ MHz.
Although the coherence time of the topological qubit may be longer, we still choose
$\gamma=1$ MHz as Hamiltonian (\ref{tp2}) is mediated by the superconducting qubit and 1
$\mu$s is much shorter than its coherence time \cite{vi}.

In summary, we have proposed  to implement tunable interfaces between the topological
qubit and the superconducting charge qubit or the  microwave cavity. Combining the two
interfaces, we are able to have control over and read out a topological qubit
as well as implement nontrivial entangling gates between two different qubits.
Therefore, the two interfaces constitute the building blocks towards universal quantum
computation with topological qubits.

\bigskip

This work was supported by the NSFC (Grants No. 11004065, No. 11125417, No. 11105136 and No.
11104096), the NFRPC (Grants No. 2013CB921804 and No. 2011CB922104), the PCSIRT, the GRF (Grants No.
HKU7058/11P and No. HKU7045/13P), the CRF (Grant No. HKU-8/11G)  of the RGC of Hong Kong,
and the URC fund of HKU.

\end{document}